\documentclass[runningheads]{llncs}
\usepackage{graphicx}
\begin{document}
%
\title{SCARLET: Explainable Attention based Graph Neural Network for Fake News spreader prediction}
%
\titlerunning{PAKDD, 21 accepted paper.}
\author{Bhavtosh Rath\inst{1} \and
Xavier Morales\inst{2}\and
Jaideep Srivastava\inst{1}}
\authorrunning{Bhavtosh Rath et al.}
\institute{University of Minnesota, USA \\
\email{rathx082@umn.edu, srivasta@umn.edu}\and
Harvard College\\
\email{xavier\_morales@college.harvard.edu}}
%
\maketitle              
\begin{abstract}
\vspace*{-.5cm}
False information and true information fact checking it, often co-exist in social networks, each competing to influence people in their spread paths.  An efficient strategy here to contain false information is to proactively identify if nodes in the spread path are likely to endorse false information (i.e. further spread it) or refutation information (thereby help contain false information spreading). 
In this paper, we propose SCARLET (truSt and Credibility bAsed gRaph neuraL nEtwork model using aTtention) to predict likely action of nodes in the spread path.
We aggregate trust and credibility features from a node's neighborhood using historical behavioral data and network structure and explain how features of a spreader's neighborhood vary. 
Using real world Twitter datasets, we show that the model is able to predict false information spreaders with an accuracy of over 87\%. 

\vspace*{-.5cm}
\end{abstract}
\vspace*{-.5cm}
\section{Introduction}
\vspace*{-.3cm} 
Social network platforms like Twitter, Facebook and Whatsapp are used by millions around the world to share information and opinions. Often, the veracity of content shared on these platforms is not confirmed. This gives rise to scenarios where information having conflicting veracity, i.e. false information and its refutation, co-exist. Refutation can be defined as true information which fact checks claims made by a false information. 
A typical scenario is that false information originates at time $t_1$, and starts propagating. Once it is identified, its refutation information is created at time $t_2$ ($t_1 < t_2$). Both pieces of information propagate simultaneously, with many nodes lying in their common spreading paths.

While detecting false information is an important and widely researched problem, an equally important problem is that of preventing the impact of false information spreading. Techniques involve containment/suppression of false information, as well as accelerating the spread of its refutation. 
{\it Being able to  predict the likely action of such users before they are exposed to false information is an important aspect of such a strategy.}  Nodes identified as vulnerable to believing false information can thus 1) be cautioned about the presence of the false information so that they do not propagate it, and 2) be urged to propagate its refutation. While optimization models based on information diffusion theories have been proposed in the past for misinformation containment, recent advancements in deep learning on graphs serve as the motivation to explore false information control models which use components that exist even before false information starts spreading, namely the underlying network structure and people's historical behavioral data.

\textit{Trust} and \textit{Credibility} are important psychological and sociological concepts respectively, that have subtle differences in their meanings. While trust represents the confidence one person has in another person, credibility represents generalized confidence in a person based on their perceived performance record~\cite{renn1991credibility}. Thus, in a graph representation of a social network, trust is a property of a (directed) edge, while credibility is a property of an individual node. Metzger et al.~\cite{metzger2013credibility} showed that \textit{the interpretation of a neighbor's credibility by a node relies on its perception of the neighbor based on their trust dynamics.} Motivated with this idea, we propose a graph neural network model that integrates people's credibility and interpersonal trust features in a social network to predict whether a node is likely to spread false information or not. We make the following contributions in this paper:\\
\textbf{1)} We propose $SCARLET$, a novel user-centric model using graph neural network with attention mechanism to predict whether a node will most likely spread false information, its refutation or be a non-spreader.\\
\textbf{2)} We demonstrate that a person's decision to spread a false information is sensitive to its perception of neighbor's credibility, and this perception is a function of trust dynamics with the neighbor.\\
\textbf{3)} To the best of our knowledge, this is the first model being evaluated on real world Twitter datasets of co-existing false and refutation information.\\
\vspace*{-.1cm}\\
 \textbf{Related Work:} Social science research in the past has explored the aspects of people's behavior that cause false information spreading. Jaeger et al. \cite{jaeger1980hears} was one of the first to study what makes rumors believable when told by peers instead of authority figures. While it focused on modelling people's anxiety, it served as motivation to explore other sociological features that are relevant to information spreading. Petty and Cacioppo \cite{petty2012communication} found credibility perception to be an important factor for believing false information. Rosnow et al. \cite{rosnow1991inside} proposed that {\it interpersonal trust} also played an important role in rumor transmission. The idea was further enforced by Morris et al. \cite{morris2012tweeting} where they claimed that people assess credibility based on trust relationships with their neighbors in a social network. 
 Motivated by these ideas, there has been much interest in computational models for false information spreader detection using trust, which has shown promising results \cite{rath2018utilizing,rath2019evaluating}.
Many computational techniques to combat false information spreading have been explored over the past decade, as summarized by Sharma et al. \cite{sharma2019combating}. Most models rely on generating relevant features from the information that help distinguish false information from true. Our proposed model is based on recent advances in graph neural networks~\cite{wu2020comprehensive}. In addition, our work proposes an explainable attention based model, inspired from recent work~\cite{lu2020gcan,shu2019defend}.
Qui et al.~\cite{qiu2018deepinf} focuses on influence in general, while our model integrates people's psychological and sociological features to identify false information spreaders.

Models inspired by information diffusion models for false information mitigation have also been proposed. Budak et al. \cite{budak2011limiting} proposed an optimization strategy to identify false information spreaders in a network who, when convinced by its refutation, would minimize the number of people receiving the false information. Nguyen et al. \cite{nguyen2012containment} proposed greedy approaches to a similar problem of limiting the spread of false information in social networks. More recently, Tong et al. \cite{tong2018misinformation} studied the problem as a multiple cascade diffusion problem. 

\vspace*{-.3cm}
\section{Interpersonal Trust and User Credibility features}
\vspace*{-.2cm}
\subsection{Trust-based features}
\vspace*{-.1cm}
\textbf{1. Global Trust ($Tr^G$) :} Global trust are  trust scores that are computed on the directed follower-followee network around information spreaders. It is called global because an individual's trust score is sensitive to changes in the network structure. Using the Trust in Social Media (TSM) algorithm~\cite{roy2016trustingness}, we quantify the likelihood of \textit{trusting others} and being \textit{trusted by others}. 
The TSM algorithm uses a directed graph $\mathcal{G(\mathcal{V},\mathcal{E})}$ as input, together with a specified convergence criteria, and computes trustingness and trustworthiness scores using the equations:
$ti(v)=\sum_{\forall x \in out(v)}\left(\frac{w(v,x)}{1+(tw(x))^s}\right)$ and $tw(u)=\sum_{\forall x \in in(u)}\left(\frac{w(x,u)}{1+(ti(x))^s}\right)$ where $u, v, x \in \mathcal{V}$ are nodes, $ti(v)$ and $tw(u)$ are the trustingness and trustworthiness scores of $v$ and $u$, respectively, $w(v,x)$ is the weight of edge from $v$ to $x$, $out(v)$ is the set of out-edges of $v$, $in(u)$ is the set of in-edges of $u$, and $s$ is the involvement score of the network. The involvement score is basically the potential risk an actor takes when creating a link in the network. Details of the algorithm are excluded due to space constraints and can be found in~\cite{roy2016trustingness}.\\
\textbf{2. Local Trust ($Tr^L$) :}  Local trust is computed based on the retweeting behavior of an individual. It is termed local because the trust score depends on node's behavior, and not on the network structure. We consider the proxy for \textit{trusting others} as the fraction of tweets of $x$ that are retweets ($RT_x$) denoted by $\sum_{\forall i \in t}\{1$ if $ i=RT_x$ else $0\}/n(t)$. Meanwhile, we consider the proxy for \textit{trusted by others} as the average number of times $x$'s tweets are retweeted ($n(RT)$) denoted by $\sum_{\forall i \in t}{i_{n(RT_x)}/n(t)}$. ($t$ represents the most recent tweets posted in $x$'s timeline).

\vspace*{-.5cm} 
\subsection{Credibility-based features}
\vspace*{-.2cm}
Credibility of users is generalized based on features extracted from information posted on their timeline, empirically studied by Castillo et al.~\cite{castillo2011information}. 
We generate relevant credibility features for nodes in the network, which can be categorized into two types: user-based and content-based.\\
\textbf{1. User-based Credibility ($Cr^U$) :} User credibility features  are extracted from user metadata of nodes in the network. Features used in our model are summarized below:

A. Registration age (U1): Registration age denotes the time that has transpired since a user created their account. Older accounts tend to be associated with more credible users.

B. Overall activity count (U2): Activity or statuses count is the number of tweets  issued by a user. Low credibility is associated with users who have less activity on their timeline.

C. Is verified (U3): This label suggests whether a user account is marked as authentic or not by Twitter. Verified accounts are more likely to be credible.\\
\textbf{2. Content-based Credibility ($Cr^C$) :} These features are obtained by aggregating a user's timeline activity. It is important to note that, unlike Castillo's assumption, we do not make a distinction between information that is specifically related to news or not, as that process would require manually assessing newsworthiness of the tweets. The following relevant features are extracted:

A. Emotions conveyed by user (M1): Emotions represent positive or negative sentiments associated with a tweet. Strong sentiments are usually associated with non-credible users.

B. Level of uncertainty (M2): Level of uncertainty is quantified as the fraction of user's tweets that are questioning in nature. Tweets with a high level of uncertainty tend to be less credible.

C. External source citation (M3): External source citation is quantified as the fraction of user's tweets that cite an external URL. Tweets which cite URLs tend to be more credible.

\vspace*{-.3cm}
\section{Proposed approach}
\vspace*{-.2cm} 
This section explains how we integrate both credibility and trust features in an attention based graph neural network model to predict whether a person would likely be a spreader of false information or its refutation. The problem formulation is as follows:\\
\textbf{Problem formulation:} Let $\mathcal{G(\mathcal{V},\mathcal{E})}$ be a directed social network containing false information spreaders ($\mathcal{V}_F$), refutation information spreaders ($\mathcal{V}_T$) and non-spreaders ($\mathcal{V}_{\hat{Sp}}$) at a time instance \textit{t} ($\{\mathcal{V}_F\cup\mathcal{V}_T \cup \mathcal{V}_{\hat{Sp}}\} \subset \mathcal{V}$). By assigning importance score using global ($Tr^G$) and local ($Tr^L$) trust features ($Tr = Tr^G || Tr^L$), and aggregating user-based ($Cr^U$) and content-based ($Cr^C$) credibility features ($Cr = Cr^U || Cr^C$) of node \textit{i}  and its neighborhood nodes ($\mathcal{N}_i{^K}$) sampled till depth \textit{K}, we predict whether \textit{i} is more likely to spread false information, refutation information or be non-spreader at future time $t + \Delta t$.

The proposed graph neural network framework can be broadly divided into two steps:

\textbf{1.} We  assign an importance score to neighborhood nodes ($\mathcal{N}_i{^K}$) sampled till depth $K$ based on trust ($Tr$) features. This is done using an attention mechanism.

\textbf{2.} We learn representations using Graph Convolutional Networks by aggregating credibility ($Cr$) features proportional to the importance scores assigned for the neighborhood nodes based on step 1.

\begin{figure*}[t]
\centering
\includegraphics[width=\linewidth]{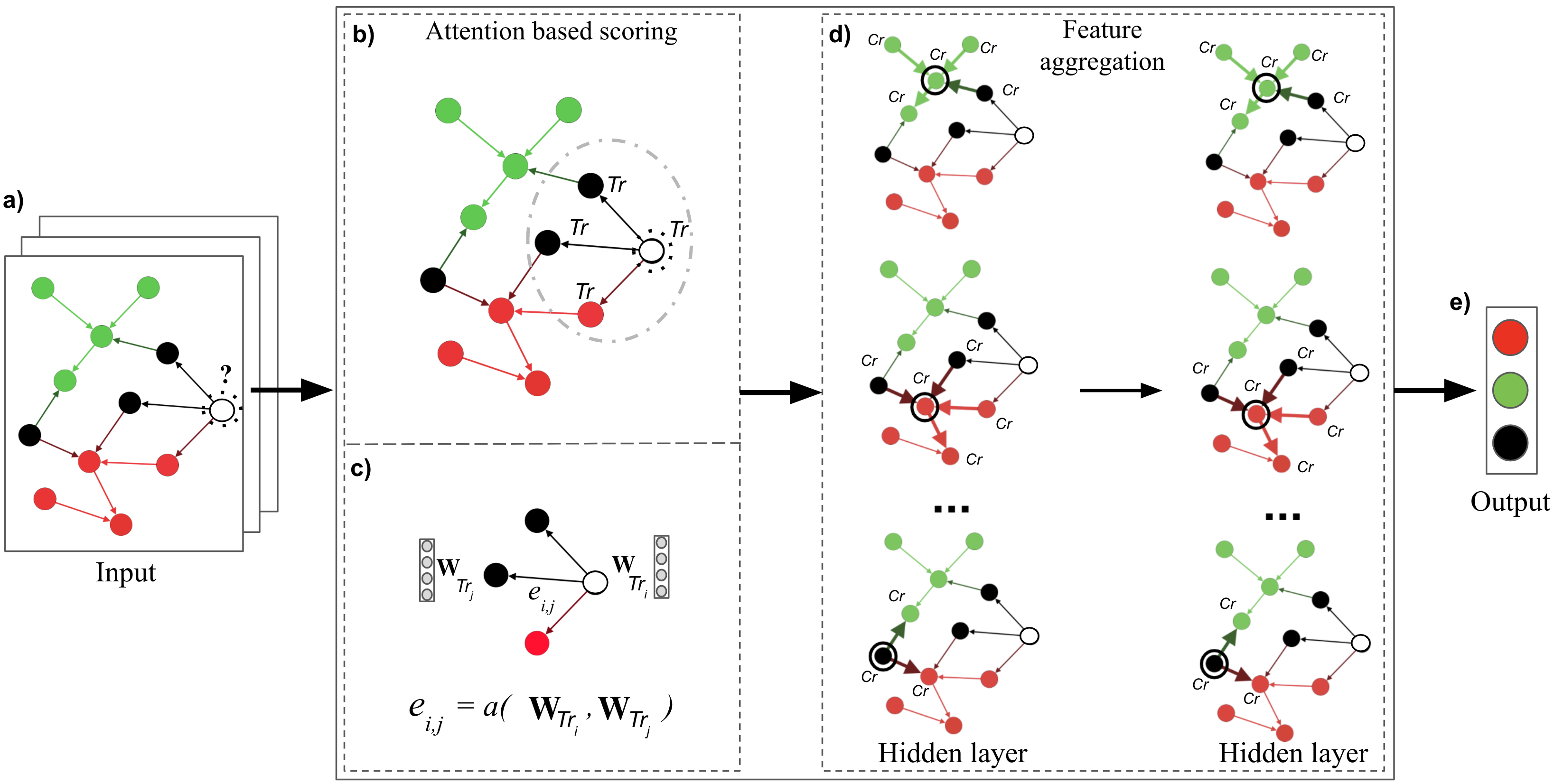}
\caption{Architecture overview. a) Node neighborhood is fed into the graph neural network.  b) Interpersonal trust dynamics is evaluated using ($Tr$) features. c) Importance score $e$ is assigned to neighbors using graph attention mechanism. d) Credibility ($Cr$) features are aggregated proportional to neighbors' importance scores using Graph Convolution Networks. e) Node is identified as either
false information spreader (red), refutation spreader (green), or a non-spreader (black).}
\label{fig:framework}
\vspace*{-.2cm} 
\end{figure*}

An overview of the proposed model architecture is shown in Figure~\ref{fig:framework}. The following subsections explain the framework in detail.\vspace*{-.1cm}\\
\vspace*{-.1cm}\\
\textbf{3.1 Importance score using attention:} We apply a graph attention mechanism \cite{velivckovic2017graph} which  attends over the neighborhood of $i$ and, based on their trust features, assigns an importance score to every $j$ ($j \in \mathcal{N}_i$). First, every node is assigned a parameterized weight matrix ($\textbf{W}$) to perform linear transformation. Then, self-attention is performed using a shared attention mechanism $a$ (a single layer feed-forward neural network) which computes trust-based importance scores. The unnormalized trust score between $i$,$j$ is represented as:
\begin{equation}
\vspace*{-.2cm}
e_{ij} = a (\textbf{W}_{Tr_i}, \textbf{W}_{Tr_j})\\
\end{equation}
where $e_{ij}$ quantifies $j$'s importance to $i$ in the context of interpersonal trust. We perform masked attention by only considering nodes in $\mathcal{N}_i$. This way we aggregate  features based only on the neighborhood's structure. To make the importance scores comparable across all neighbors we normalize them using the softmax function:
\begin{equation}
\vspace*{-.2cm}
\alpha_{ij} = softmax(e_{ij}) = \frac{exp(e_{ij})}{\sum_{ k \in \mathcal{N}_i}exp(e_{ik})}\\
\end{equation}
The attention layer $a$ is parameterized by weight vector $\vec{\textbf{a}}$ and applied using LeakyReLU nonlinearity. Normalized neighborhood edge weights can be represented as:
\begin{equation}
\vspace*{-.2cm}
\alpha_{ij} =  \frac{exp(LeakyReLU(\vec{\textbf{a}}^{T}[\textbf{W}_{Tr_i} || \textbf{W}_{Tr_j}]))
}
{\sum_{ k \in \mathcal{N}_i}
exp(LeakyReLU
(\vec{\textbf{a}}^{T}[\textbf{W}_{Tr_i} 
|| 
\textbf{W}_{Tr_k}
]))}
\end{equation}
 $\alpha_{ij}$ thus represents trust between $i$ and $j$ with respect to all nodes in $\mathcal{N}_i$.  Each $\alpha_{ij}$ obtained for the edges is used to create an attention-based adjacency matrix $\hat{A}_{atn} = [\alpha_{ij}]_{|\mathcal{V}| \times |\mathcal{V}|}$ which is later used to aggregate credibility features.\vspace*{.3cm} \\
\textbf{3.2 Feature aggregation:} The Graph Convolution Network~\cite{kipf2016semi} is a graph neural network model that efficiently aggregates features from a node's neighborhood. It consists of multiple neural network layers where the information propagation between layers can be generalized by Equation~\ref{eq:5_featAg}. Here, $H$ represents the hidden layer and $A$ represents the adjacency matrix representation of the subgraph ($A = \hat{A}_{atn}\prime $). $H^{(0)}= Cr$ and $H^{(L)}= Z$, where $Z$ denotes node-level output during transformation.
\begin{equation}
\vspace*{-.3cm} 
H^{(l+1)}=f(H^{(l)}, A) \\
\label{eq:5_featAg}
\end{equation}
We implement a Graph Convolution Network with two hidden layers using a propagation rule as explained in \cite{kipf2016semi}.
\label{eq:2_1}
\begin{equation}
\vspace*{-.3cm}
H^{(l+1)}= \sigma(\hat{D}^{-1/2}\hat{A}\hat{D}^{-1/2}H^{(l)}W^{(l)})\\
\end{equation}
Here, $\hat{A}= A + I$, where $I$ is the identity matrix of the neighborhood subgraph. This operation ensures that we include self-features during aggregation of neighbor's credibility features. $\hat{D}$ is the diagonal matrix of node degrees for $\hat{A}$, where $\hat{D}_{ii} = \sum_j \hat{A}_{ij}$. $W^{(l)}$ is the layer weight matrix, and $\sigma$ denotes the activation function. Symmetric normalization of $\hat{D}$ ensures our model is not sensitive to varying scale of the features being aggregated.\\
\vspace*{-.1cm}\\
\textbf{3.3 Node classification:} Using credibility features and network structure for nodes in $i$'s neighborhood, node representations are learned from the graph using a symmetric adjacency matrix with attention-based edge weights ($\hat{A} = \hat{D}^{-1/2}\hat{A}_{atn}\prime\hat{D}^{-1/2}$). Following forward propagation model is applied:
\begin{equation}
\vspace*{-.2cm}
Z = f(X, \hat{A}_{atn}\prime) = 
softmax(\hat{A} ReLU(\hat{A}X{W^{(0)}})W^{(1)})\\
\end{equation}
$X$ represents the credibility features.  $W^{(0)}$ and $W^{(1)}$ are input-to-hidden and hidden-to-output weight matrices respectively, and are learnt using gradient descent learning. Classification is performed using the following cross entropy loss function:
\begin{equation}
\vspace*{-.2cm}
\mathcal{L} = {\sum_{l \in \mathcal{Y}_L}\sum_{f \in Cr}}Y_{lf}lnZ_{lf}\\
\end{equation}
where $\mathcal{Y}_L$ represents indices of labeled vertices, $f$ represents each of the credibilty features being used in the model, and $Y \in R^{|\mathcal{Y}_L| \times |Cr|}$ is the label indicator matrix.\vspace*{-.5cm}\\
\vspace*{-.3cm} 
\begin{table*}[h]
        \caption{Network dataset statistics for news events N1-N10.}
    \label{tab:4_1}
        \resizebox{\textwidth}{!}{%
    \begin{tabular}{|c|c|c|c|c|c|c|c|c|c|c|c|c|c|c|c|} 
    \hline
    \textbf{} &
    \multicolumn{3}{c|}{\textbf{N1}} & 
    \multicolumn{3}{c|}{\textbf{N2}} &
    \multicolumn{3}{c|}{\textbf{N3}} &
    \multicolumn{3}{c|}{\textbf{N4}} &
    \multicolumn{3}{c|}{\textbf{N5}}\\\cline{2-16}
    \textbf{} & \textbf{$|\mathcal{V}|$} & \textbf{$|\mathcal{E}|$} & $\vert Sp \vert$ & 
    \textbf{$|\mathcal{V}|$} & \textbf{$|\mathcal{E}|$} & $\vert Sp \vert$ & 
    \textbf{$|\mathcal{V}|$} & \textbf{$|\mathcal{E}|$} & $\vert Sp \vert$ &
    \textbf{$|\mathcal{V}|$} & \textbf{$|\mathcal{E}|$} & $\vert Sp \vert$ &
    \textbf{$|\mathcal{V}|$} & \textbf{$|\mathcal{E}|$} & $\vert Sp \vert$ \\
    \hline
    \textbf{F} & 1,797,059 & 5,316,114 & 2,584 & 885,598 & 1,824,585 & 943 & 1,228,479 & 2,477,986  & 1,313 & 2,607,629 & 7,146,454 & 4,552 & 2,150,820 & 5,215,120 & 3,344\\
    \textbf{T} & 1,164,162 & 2,283,160 & 437 & 453,537 & 879,854 & 403 & 1,169,681 & 1,988,576 & 425 & 433,616 & 773,778 & 467 & 1,168,820 & 1,543,513 & 305 \\
    \textbf{F $\cup$ T} & 2,677,924 & 7,562,503 & 3,017 & 1,230,559 & 2,641,513 & 1,337 & 2,198,524 &  4,458,228 & 1,738 & 2,900,925 & 7,882,019 & 5,015 & 3,019,066 & 6,631,032 & 3,627\\ 
    \textbf{F $\cap$ T} & 283,297 & 8,956 & 4 & 108,576 & 59,912 & 9 & 199,636 & 376 & 0 & 140,320 & 3,273 & 5 & 300,574 & 112,098 & 22\\\hline
    \textbf{} &
    \multicolumn{3}{c|}{\textbf{N6}} & 
    \multicolumn{3}{c|}{\textbf{N7}} &
    \multicolumn{3}{c|}{\textbf{N8}} &
    \multicolumn{3}{c|}{\textbf{N9}} &
    \multicolumn{3}{c|}{\textbf{N10}}\\\cline{2-16}
    \textbf{} & \textbf{$|\mathcal{V}|$} & \textbf{$|\mathcal{E}|$} & $\vert Sp \vert$ & 
    \textbf{$|\mathcal{V}|$} & \textbf{$|\mathcal{E}|$} & $\vert Sp \vert$ & 
    \textbf{$|\mathcal{V}|$} & \textbf{$|\mathcal{E}|$} & $\vert Sp \vert$ &
    \textbf{$|\mathcal{V}|$} & \textbf{$|\mathcal{E}|$} & $\vert Sp \vert$ &
    \textbf{$|\mathcal{V}|$} & \textbf{$|\mathcal{E}|$} & $\vert Sp \vert$ \\
    \hline
    \textbf{F} & 2,387,610 & 5,356,288 & 3,498 & 627,147 & 1,071,120 & 696 & 2,036,162 & 2,876,783 & 894 & 1,197,935 & 2,139,912 & 2,317 & 2,174,023 & 4,280,962 & 2,323\\
    \textbf{T} & 1,297,371 & 1,727,503 & 481 & 1,166,528 & 2,524,907 & 847 & 1,058,482 & 1,513,404  & 489 & 2,999,865 & 6,317,032 & 1,833 & 704,006 & 1,314,996 & 741\\
    \textbf{F $\cup$ T} & 2,449,434 & 5,691,728 & 3,769 & 1,606,924 & 3,577,449 & 1,534 & 2,663,392 & 4,082,373 & 1,365 & 4,064,545 & 8,443,888 & 4,151 & 2,729,312 & 5,584,915 & 3,063\\
    \textbf{F $\cap$ T} & 1,235,547 & 1,379,510 & 212 & 186,751 & 11,131 & 9 & 431,252 & 305,358  & 20 & 133,255 & 722 & 1 & 148,717 & 699 & 1 \\ \hline
   \end{tabular}
   \vspace*{-.3cm} 
    }
\vspace*{-.3cm} 
\end{table*}
\vspace*{-.5cm} 
\vspace*{-.4cm}
\section{Experimental Analysis}
\begin{figure*}[h]
\vspace*{-.7cm}
\centering
\includegraphics[width=\linewidth]{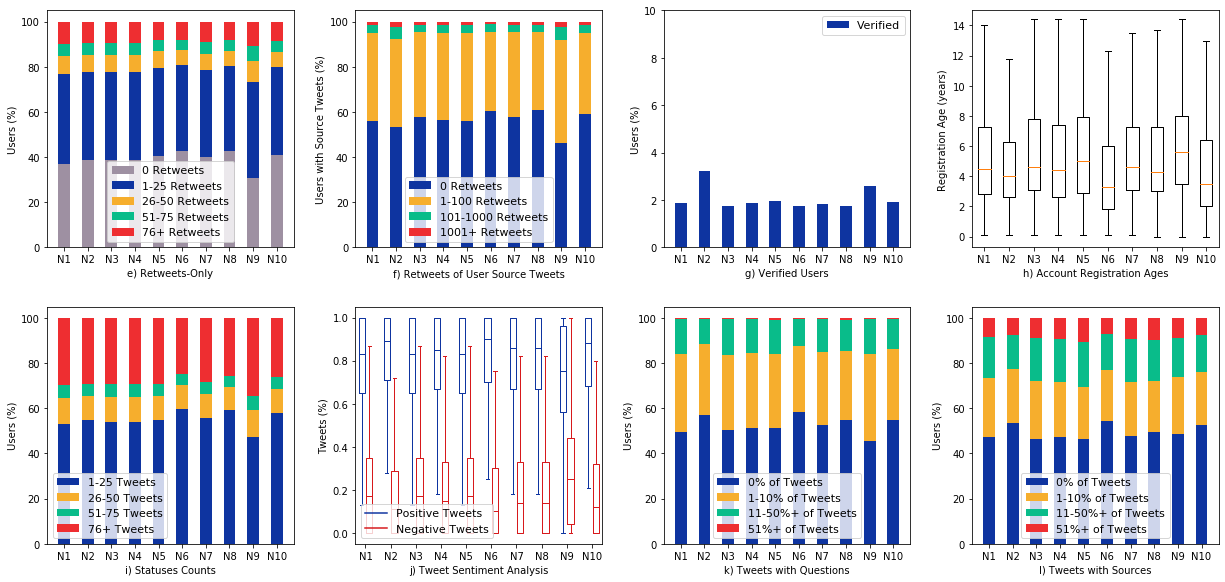}
\caption{Trust and credibility feature analysis from networks N1-N10.}
\label{fig:complete_features}
\vspace*{-.5cm}
\end{figure*}

\begin{figure}[h]
\vspace*{-.4cm} 
\centering
\includegraphics[width=.6\linewidth]{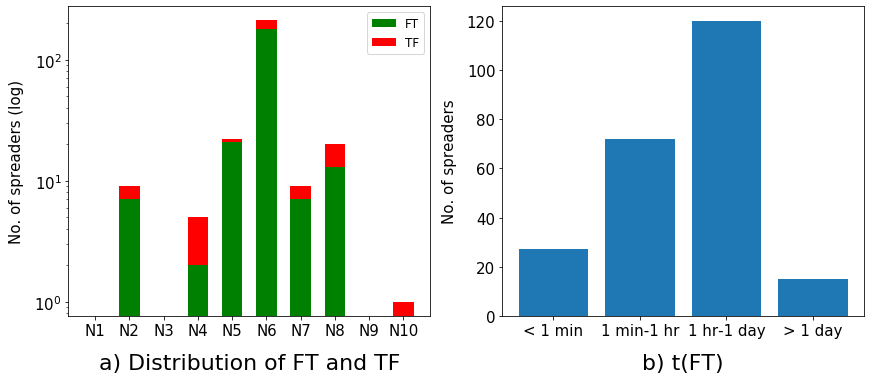}
\caption{Analysis of spreaders in F $\cap$ T.}
\label{fig:FT_analysis}
\vspace*{-.5cm} 
\end{figure}

\vspace*{-.1cm}
\textbf{4.1 Data collection:} We evaluate our proposed model using real world Twitter datasets. The ground truth of false information and the refuting true information was obtained from \textit{www.altnews.in}, a popular fact checking website based in India. The source tweet related to the information was obtained directly as a tweet embedded in the website.
From that source tweet, we used the Twitter API to determine the source tweeter and retweeters (proxy for spreaders), the follower-following network of the spreaders (proxy for social network), and user activity data (100 most recent tweets) for all nodes in the network. Trust and credibility features extracted from the activity data are summarized in Figure~\ref{fig:complete_features}. Besides evaluating our model on the false information (F) and true information (T) spreading networks separately, we also evaluated our model on the combined information spreading networks (F $\cup$ T). Details regarding the number of nodes ($|\mathcal{V}|$), edges ($|\mathcal{E}|$), and spreaders ($\vert Sp \vert$) for the networks of 10 different news events (N1-N10) is detailed in Table \ref{tab:4_1}.\vspace*{-.2cm}\\ 
\\
\textbf{4.2 Analysis of F $\cap$ T:} F $\cap$ T in Table \ref{tab:4_1} denotes the section of the network that was exposed to both the false and its refutation information. An interesting observation is the spreaders who decided to spread both types of information. Figure \ref{fig:FT_analysis} (a) denotes the distribution of spreaders in F $\cap$ T who spread false information followed by its refutation (FT) and  those whose spread refutation followed by the false information (TF). N1 and N9 is excluded from the analysis as our dataset as we did not have the spreaders' timestamp information. An interesting observation is that the majority of spreaders belong to FT. Intuitively, these are spreaders who trusted the endorser without verifying the information and later corrected their position, thereby implying that they did not intentionally want to spread false information. Consequently, the proposed model can help identify such people proactively in order to take measures to prevent them from endorsing false information in the first place. While spreaders belonging to TF are comparatively fewer (whose intentions are not certain) the proposed model can help identify them and effective containment strategies can be adopted. Figure \ref{fig:FT_analysis} (b) shows the time that transpired between spreading refutation and false information for FT spreaders. Once the false information is endorsed, large portions of the network must have already been exposed to false information before the endorser corrected themselves after a significant amount of time ($\sim$ 1 day). This serves as a strong motivation to have a spreader prediction model which proactively identifies likely future spreaders.\vspace*{-.5cm} \\
\\
\textbf{4.3 Models and metrics:} We compare our proposed attention based model with 10 baseline models. Among the baselines, 3 models use node features only ($SVM_{Tr}$, $SVM_{Cr}$, $SVM_{Tr, Cr}$), 1 model uses network structure only ($LINE$) and 6 models integrate both node features and the network structure ( $SAGE_{Tr}$, $SAGE_{Cr}$, $SAGE_{Tr, Cr}$, $GCN_{Tr}$, $GCN_{Cr}$, $GCN_{Tr, Cr}$). 
\\
\textbf{1. Node feature-based models:}\\
i). $SVM_{Tr}$: This model applies Support Vector Machines (SVM)~\cite{cortes1995support} on node's trust based features $Tr$ to find an optimal classification threshold.\\
ii). $SVM_{Cr}$: This model applies SVM on node's credibility based features $Cr$.\\
iii). $SVM_{Tr, Cr}$: This model applies SVM by combining node's trust based and credibility based features.\\
\textbf{2. Network structure-based models:}\\
iv). $LINE$: Applies the Large-scale Information Network Embedding~\cite{tang2015line} as a transductive representation learning baseline, where node embeddings are generated after optimization is performed on the entire graph structure.\\
\textbf{3. Network structure + node feature-based models:}\\
v). $SAGE_{Tr}$: GraphSAGE~\cite{hamilton2017inductive} serves as the inductive learning baseline where node embeddings are generated by aggregating $Tr$ features from neighborhoods.\\
vi). $SAGE_{Cr}$: This inductive representation learning baseline generates node embeddings by aggregating $Cr$ features from neighborhoods.\\
vii). $SAGE_{Tr, Cr}$: This inductive representation learning baseline generates node embeddings by aggregating both $Tr$ and $Cr$ features from   neighborhoods.\\
viii). $GCN_{Tr}$: This model applies Graph Convolution Networks~\cite{kipf2016semi} to learn node embeddings by aggregating $Tr$ features from neighborhoods.\\
ix). $GCN_{Cr}$: This model applies Graph Convolution Networks by aggregating $Cr$ features from  neighborhoods.\\
x). $GCN_{Tr, Cr}$: This model applies Graph Convolution Networks  by aggregating both $Tr$ and $Cr$ features from  neighborhoods.\vspace*{-.3cm}\\

\begin{figure}[h]
\vspace*{-.5cm} 
\centering
\hspace*{-.5cm} 
\includegraphics[width=\linewidth]{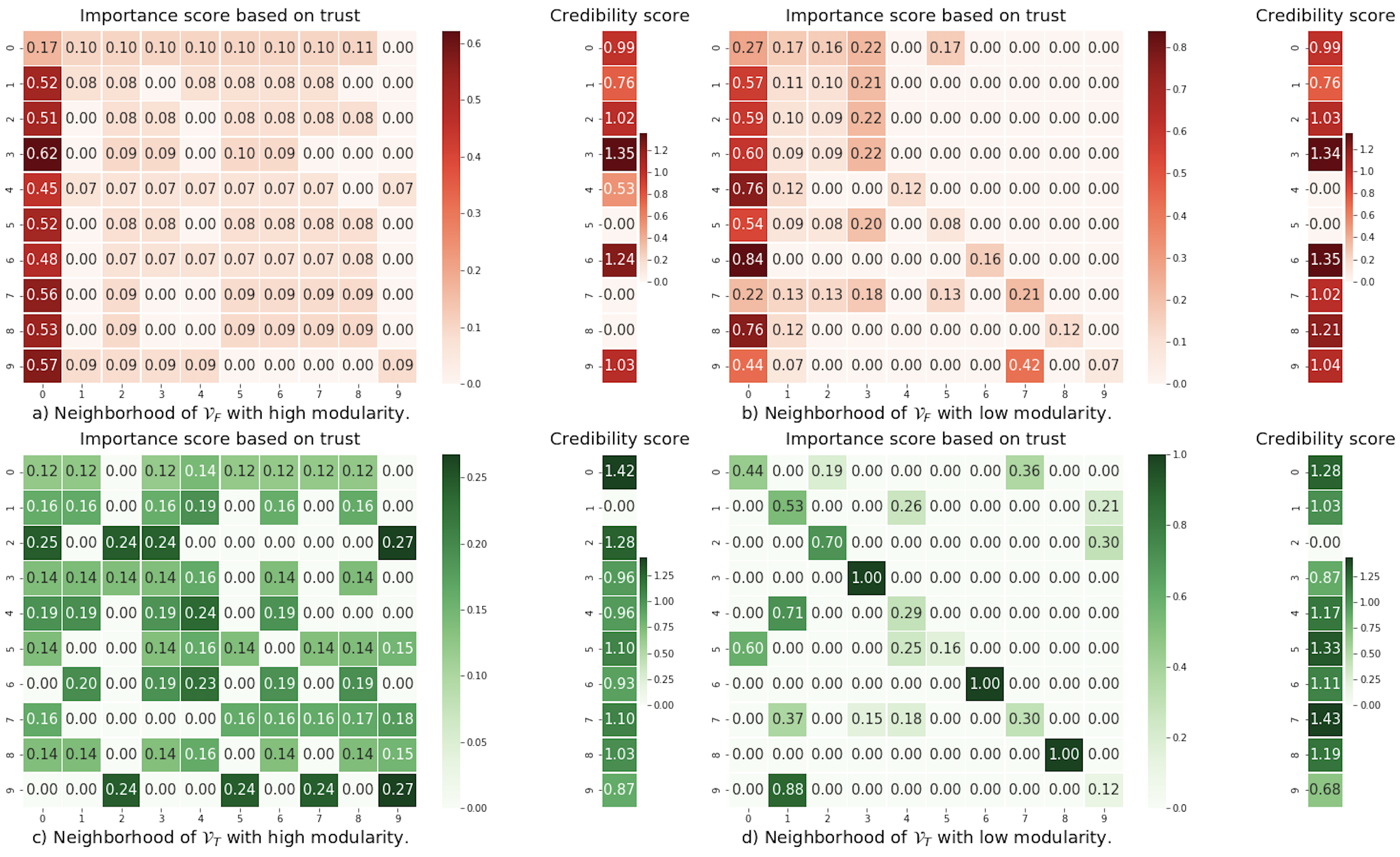}
\caption{Explaining trust and credibility of spreader's neighbors.}
\label{fig:exp_analysis}
\vspace*{-.4cm} 
\end{figure}

$SCARLET$ is the proposed model in this paper, which aggregates a node neighborhood's $Cr$ features based on attention based importance scores assigned using $Tr$. For evaluation, we did an 80-10-10 train-validation-test split of the dataset. We used 5-fold cross validation and four common metrics: Accuracy, Precision, Recall, and F1 score. \\
\vspace*{-.3cm} 
\\
\textbf{4.4 Implementation details:} We obtained Global Trust features by running the TSM algorithm on the follower-following network of the spreaders. We used the generic settings for TSM parameters (number of iterations = 100, involvement score = 0.391) based on~\cite{roy2016trustingness}. 
The size of sampled neighborhood was set to 50 and depth was set to 1. We considered neighbors with higher degrees in order to generate denser adjacency matrices.  The number of epochs, batch size, learning rate and dropout rate were set to 200, 64, 0.001 and 0.2, respectively.
The code implementation is also available\footnote{https://github.com/BhavtoshRath/GAT-GCN-SpreaderPrediction}.\vspace*{-.3cm}\\
\vspace*{-.3cm}
\begin{table*}[h]
\vspace*{-.2cm}
        \caption{Model performance evaluation ($\mathcal{V}_F$): False information spreader, ($\mathcal{V}_T$): Refutation spreader.}
    \label{tab:spreader_pred}
        \resizebox{\textwidth}{!}{%
    \begin{tabular}{|l|c|c|c|c|c|c|c|c|c|c|c|c|} 
    \hline
\textbf{} &
    \multicolumn{4}{c|}{\textbf{F ($\mathcal{V}_F$)}} & 
    \multicolumn{4}{c|}{\textbf{T ($\mathcal{V}_T$)}} &
    \multicolumn{4}{c|}{\textbf{F $\cup$ T ($\mathcal{V}_F$)}}\\\cline{2-13}
     \textbf{} & \textbf{Accu.} & \textbf{Prec.} & \textbf{Rec.} & 
    \textbf{F1} & \textbf{Accu.} & \textbf{Prec.} & \textbf{Rec.} & 
    \textbf{F1} & \textbf{Accu.} & \textbf{Prec.} & \textbf{Rec.} & 
    \textbf{F1} \\
    \hline
    $SVM_{Tr}$ & 0.497 & 0.512 & 0.468 & 0.478 & 0.473 & 0.472 & 0.452 & 0.445 & 0.398 & 0.19 & 0.465 & 0.229\\
    $SVM_{Cr}$ & 0.508 & 0.517 & 0.517 & 0.509 & 0.501 & 0.477 & 0.565 & 0.509 & 0.408 & 0.196 & 0.542 & 0.272 \\
    $SVM_{Tr, Cr}$ &  0.516 & 0.514 & 0.579 & 0.53 & 0.52 & 0.513 & 0.598 & 0.545  & 0.444 & 0.193 & 0.489 & 0.267 \\
    $LINE$ & 0.686 & 0.626 & 0.896 & 0.733 & 0.635 & 0.608 & 0.881 & 0.717 & 0.688 & 0.71 & 0.896 & 0.786  \\
    $SAGE_{Tr}$ & 0.734 & 0.762 & 0.691 & 0.722 & 0.680 & 0.698 & 0.719 & 0.705 & 0.752 & 0.743 & 0.859 & 0.793\\
    $SAGE_{Cr}$ & 0.747 & 0.772 & 0.710 & 0.736 & 0.714 & 0.692 & 0.764 & 0.725 & 0.764 & 0.747 & 0.881 & 0.805\\
    $SAGE_{Tr, Cr}$ & 0.779 & 0.831 & 0.720 & 0.763 & \textbf{0.755} & \textbf{ 0.787} & 0.732 & 0.755 & 0.785 & 0.764 & 0.878 & 0.814\\
    $GCN_{Tr}$ & 0.784 & 0.726 & 0.947 & 0.821 & 0.718 & 0.675 & 0.916 & 0.767 & 0.753 & 0.783 & 0.930 & 0.845\\
 $GCN_{Cr}$ & 0.800 & 0.742 & 0.953 & 0.834 & 0.731 & 0.697 & 0.906 & 0.773 & 0.762 & 0.786 & 0.940 & 0.851\\
  $GCN_{Tr, Cr}$ & 0.824 & 0.774 & 0.942 & 0.848 & 0.743 &  0.702 & 0.916 & 0.783 & 0.776 & \textbf{0.788} & 0.954 & 0.861\\
    $SCARLET$& \textbf{0.876} & \textbf{0.834} & \textbf{0.966} & \textbf{0.893} & 0.734 & 0.674 & \textbf{0.981} & \textbf{0.794} & \textbf{0.789} & 0.785 & \textbf{0.972} & \textbf{0.866}\\
    \hline
  \end{tabular}
    }
\vspace*{-.5cm} 
\end{table*}
\\
\textbf{4.5 Performance evaluation:} Classification results of the baselines and proposed model are summarized in Table~\ref{tab:spreader_pred}. The results are averaged over the 10 news events. We report the precision, recall, and F1 scores of the false information spreaders class ($\mathcal{V}_F$) in F and F $\cup$ T networks, and of the refutation spreaders class ($\mathcal{V}_T$) in T network.
Due to class imbalance, we undersample the majority class to obtain balanced class distribution. We observe that structure only baseline performs better than feature only baselines, and models that combine both node features and network structure show further improvement in performance. Additionally, we observe that $Cr$ features perform better than $Tr$ features (because there are more number of $Cr$ features than $Tr$ features) and the model performance increases when we use $Tr$ and $Cr$ features together. 
$LINE$, the structure only baseline, performs better than feature only baselines by a substantial margin, which suggests that network structure plays an important role in identifying false information spreaders. In terms of accuracy, the $LINE$ model shows an increase of 32.9\%, 22.1\% and 54.9\% for F, T and F $\cup$ T networks, respectively, over $SVM_{Tr, Cr}$. Graph neural network  baselines that combine both network structure and node features show a significant improvement in performance. 
$GCN$ models perform better than $GraphSAGE$ models on all metrics for F networks, while that is not the case for T and F $\cup$ T networks. This is because $Tr$ and $Cr$ features for neighborhood of refutation information spreaders and non-spreaders do not differ much from each other.
Our proposed model $SCARLET$ shows an increase in performance for all three networks. However, $SAGE_{Tr, Cr}$ shows better accuracy and precision on T networks because the specific news events on which it performed better involved religious tones, and so decision to refute them is more sensitive to neighborhood's $Cr$ than $Tr$. Precision on F $\cup$ T networks is highest for $GCN_{Tr, Cr}$, though it is still comparable to the proposed model's performance. More importantly, in the F $\cup$ T network we observe highest accuracy and F1 scores of 78.9\% and 86.6\% , thus supporting our hypothesis that false information spreading is very sensitive to trust and credibility.\vspace*{-.4cm} \\
\begin{figure}[h]
\vspace*{-.5cm} 
\centering
\includegraphics[width=.9\linewidth]{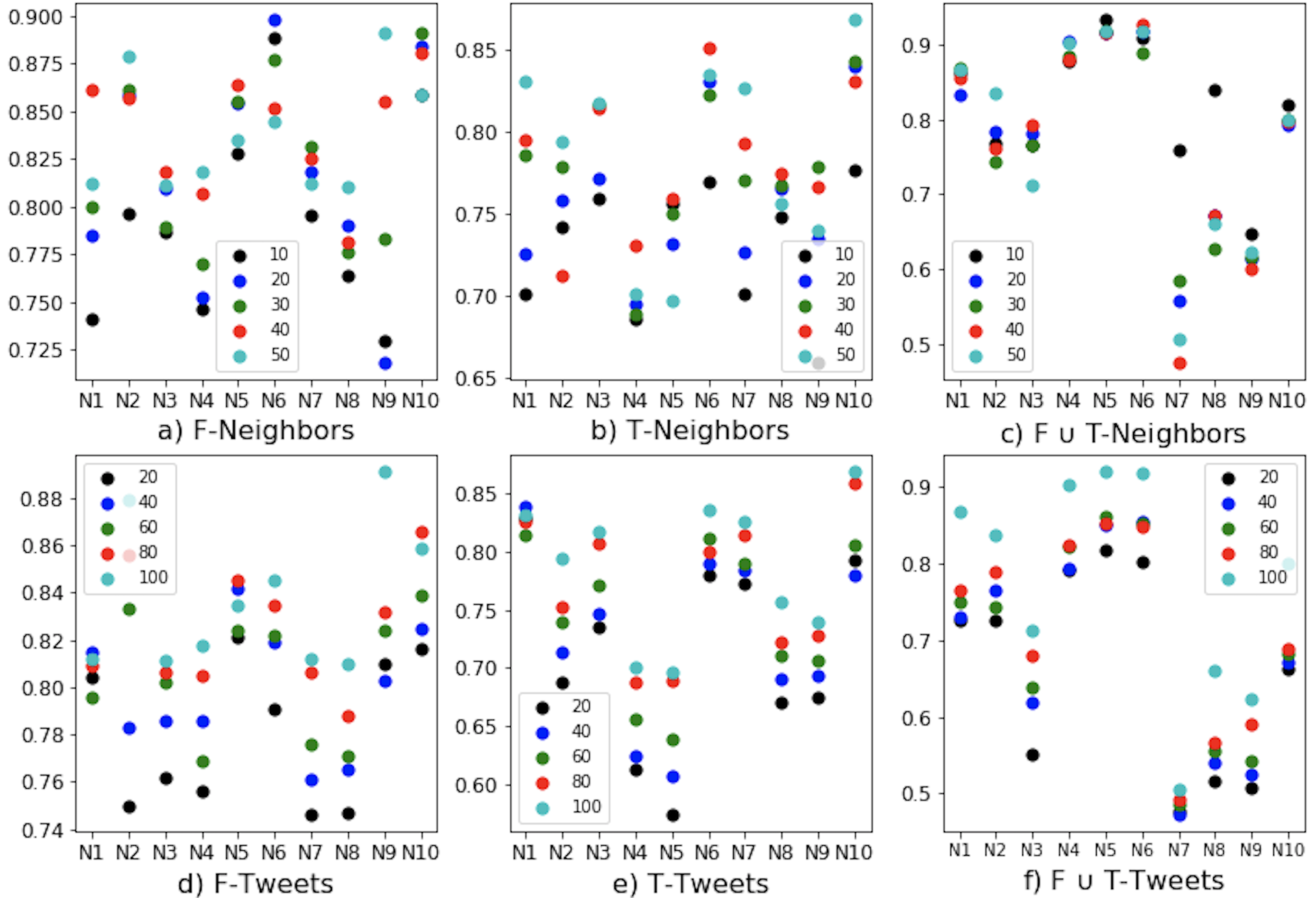}
\vspace*{-.3cm} 
\caption{Sensitivity analysis: Neighborhood size (Neighbors) and features (Tweets).}
\label{fig:param_analysis}
\vspace*{-.5cm} 
\end{figure}
\\
\textbf{4.6 Sensitivity analysis:} Figure~\ref{fig:param_analysis} shows the sensitivity analysis of F1 scores of the proposed model on two important parameters: the size of neighborhoods (Neighbors), and the number of recent tweets from user timeline (Tweets).\\
\textbf{Neighbors:} We evaluated our model on n-neighbors, where n = 10, 20, 30, 40, 50. Figure~\ref{fig:param_analysis} a), b), and c) show results on F, T and F $\cup$ T networks, respectively. We observe that model performance is not very sensitive to varying neighborhood size, which could be attributed to the fact that since we have only the immediate follower-following network (sampling depth=1) we are not able to entirely capture meaningful dynamics (i.e. the decision to retweet might depend less on the immediate neighbors, and more on the source tweeter).\\
\textbf{Tweets:} We also evaluated our model on the n-most recent timeline tweets, where n = 20, 40, 60, 80, 100. Figure~\ref{fig:param_analysis} d), e), and f) shows results on F, T and F $\cup$ T networks, respectively. We observe that for all three networks, prediction performance tends to increase as the number of timeline tweets used to aggregate features increases. This is probably because using more behavioral data helps us estimate trust and credibility features better.\vspace*{.2cm}\\
\textbf{4.7 Explainability analysis of trust and credibility:} Figure~\ref{fig:exp_analysis} shows importance scores that false ($\mathcal{V}_F$) and refutation ($\mathcal{V}_T$) spreader's neighbors (size=10) assign each other based on trust dynamics (softmax attention score) and credibility score (euclidean norm of normalized feature vector) for neighbors with both high and low modularity. Node 0 is the neighbor that the spreader endorses. We observe that $\mathcal{V}_T$'s neighbors have higher credibility than $\mathcal{V}_F$'s neighbors because of network homophily. Also low magnitude of importance scores for neighbors of node 0 of $\mathcal{V}_F$ suggest that it's neighbors trust each other less compared to $\mathcal{V}_T$'s neighbors.  We observe in Figure~\ref{fig:exp_analysis} a) and b) that node 0 in $\mathcal{V}_F$'s neighbor has strong trust dynamics with its followers (i.e. incoming edges) because it has more incoming edges than outgoing edges and also retweets and gets retweeted substantially more by the neighbors, unlike who $\mathcal{V}_T$ endorses in Figure~\ref{fig:exp_analysis} c) and d), because $\mathcal{V}_T$'s decision to endorse depends more on information source, which is usually a fact checker.

\vspace*{-.7cm}
\section{Conclusions and Future work}
\vspace*{-.3cm}
We propose $SCARLET$, an attention-based explainable graph neural network model to predict whether a node  is likely to spread false information or not. The model learns node embeddings by first assigning trust-based importance scores and then aggregating its neighborhood's credibility features proportionally. What makes this model different from most existing research is that it does not rely on features extracted from the information itself. Thus it can be used to predict spreaders even before information spreading  begins. As part of future work, we would like to analyze our model on more news events comprising larger networks in order to sample and aggregate features at greater sampling depths. 

\end{document}